# Microstructural Evolution of Charged Defects in the Fatigue Process of Polycrystalline BiFeO$_3$ Thin Films

Qingqing Ke,[a] Amit Kumar,[b] Xiaojie Lou,[c] Yuan Ping Feng,[d] Kaiyang Zeng,[b] Yongqing Cai [d,e] * and John Wang [a,]*

[a] Department of Materials Science and Engineering, National University of Singapore, Singapore 117574, Singapore
[b] Department of Mechanical Engineering, National University of Singapore, Singapore 117576, Singapore
[c] Frontier Institute of Science and Technology, Xi'an Jiaotong University, Xi'an 710049, China
[d] Department of Physics, National University of Singapore, Singapore 117542, Singapore
[e] Institute of High Performance Computing, 1 Fusionopolis Way, Singapore 138632, Singapore

**Abstract:** Fatigue failure in ferroelectrics has been intensively investigated in the past few decades. Most of the mechanisms discussed for ferroelectric fatigue have been built on the "hypothesis of variation in charged defects", which however are rarely evidenced by experimental observation. Here, using a combination of complex impedance spectra techniques, piezoresponse force microscopy and first-principles theory, we examine the microscopic evolution and redistribution of charged defects during the electrical cycling in BiFeO$_3$ thin films. The dynamic formation and melting behaviors of oxygen vacancy ($V_O$) order are identified during the fatigue process. It reveals that the isolated $V_O$ tend to self-order along grain boundaries to form a planar-aligned structure, which blocks the domain reversals. Upon further electrical cycling, migration of $V_O$ within vacancy clusters is accommodated with a lower energy barrier (~0.2 eV) and facilitates the formation of nearby-electrode layer incorporated with highly concentrated $V_O$. The interplay between the macroscopic fatigue and microscopic evolution of charged defects clearly demonstrates the role of ordered $V_O$ cluster in the fatigue failure of BiFeO$_3$ thin films.

*Supporting Information Placeholder*

## INTRODUCTION

Ferroelectric fatigue defined as a suppression of switchable polarization upon repetitive electrical cycling is a major problem in ferroelectric nonvolatile memories through degrading their life time and performance. Extensive theoretical considerations have been forwarded to account for this issue, including electromigration of oxygen vacancy ($V_O$) to form two-dimensional defects capable of pinning domains,[1] generation of a "dead or blocking layer",[2] domain blocking with electronic charges,[3] and interface nucleation inhibitions.[4] The polarization degradation was postulated to result from variation in charged defects,[5] which generally disputes on the roles of $V_O$ and injected electrons.[6] One well-known model built upon "$V_O$ hypothesis" suggests that $V_O$ blocks the domain reversal and degrades the remnant polarization through creating electroneutral complexes: "wall + compensating charge".[3] There is also a belief that the injected electrons from electrodes promote producing charged defects giving rise to fatigue failure by forming non-switching layers, local imprint or pinning the domain walls.[7-9] To properly understand the nature of the fatigue failure, there is a need to monitor the evolution of the fatigue-induced defects, which helps to develop fatigue-free ferroelectric capacitors for nonvolatile memory applications.

Bismuth ferrite (BiFeO$_3$, denoted as BFO), as a multiferroic perovskite possessing both ferroelectric and antiferromagnetic orders at room temperature, promises great potential applications in magnetoelectric and ferroelectric devices.[10-12] However, the serious fatigue failure degrades its reliability, especially largely hinders the potential application in multifunctional field through losing the controllability of magnetic order switching.[13] Great attempts have been made to improve the fatigue endurance of BFO through chemical doping.[14-16] Few works, however, have been dedicated to identify the exact mechanism of fatigue failure in BFO thin films, which therefore still remains largely obscure. For example, an orientation-dependent fatigue behavior has been reported by Baek *et al.*, who attributed the fatigue failure of single-domain BFO thin films to local domain wall pinning by charged defects.[17] Balke *et al.,* however, proposed a different mechanism that is related to the space-charge-assisted blocking of domain nucleation nearby the electrode through imaging the in-plane domain evolution by piezoresponse force microscopy (PFM) method.[18] While these works conjecture the redistribution of space charges, direct evidence related to the variation of defects structure during fatigue process has not been presented. Moreover, previous studies mainly focus on epitaxial films, neglecting the role of grain boundaries, which are supposed to significantly affect the domain dynamics through domain-wall pinning.[19] Indeed, polycrystalline BFO thin films are required in potential industrial applications, and it would be highly desirable to understand the fatigue nature of the BFO polycrystalline thin films.[20] In this context, we have investigated the interaction between polarization fatigue and evolution of space charges by conducting temperature-dependent impedance spectra and PFM studies on polycrystalline BFO thin films. First-principles calculations are also performed to explore the stability of $V_O$ clusters in BFO thin films.

### RESULTS AND DISCUSSION

**Polarization Switching in BFO.** Figure 1a shows the normalized polarization switching of BFO thin film under a bipolar triangle pulse with an amplitude of 600 kV/cm at 100 kHz. A loss of polarization about 90% is observed which occurs after $10^5$ cycles and remains nearly stable thereafter. The degraded sample (fatigued for $2\times10^9$ cycles) is found to be completely restored to the fresh state with an enhanced polarization through applying a high electric field (e.g., 1100 kV/cm) as indicated in inset of Figure 1a. This rejuvenated behavior in polarization thus excludes the effect of any permanent damages on the degradation, such as microcracks[21,22] or nonferroelectric phase (e.g., pyrochlorelike phase in PZT[23]), which leads to a reduced polarization being unable to be restored by electric stress.

applied field for fresh sample and the sample fatigued after $2\times10^9$ cycles. The jumping stages of I and II are highlighted in pink and blue colors, respectively.

To further examine the characteristics of ferroelectric behavior in the fatigued samples, field-dependent coercive field ($E_c$) and remnant polarization ($P_r$) for the fresh and fatigued samples are monitored and shown in Figure 1b. For the fresh sample, $E_c$ and $P_r$ increase rapidly near the field around 460-560 kV/cm and saturate above 560 kV/cm. Compared to the fresh sample, a couple of new features are observed in the fatigued one: (i) two jumping stages (I and II) are observed in the $E_c$ and $P_r$ curves, which are believed to relate to the reorientation of randomly-distributed domains; (ii) $E_c$ and $P_r$ in the fatigued sample exhibit lower values below 1100 kV/cm, but exceed those of the fresh sample beyond stage II. It reveals that a higher field is needed to realize the domain reversal in the fatigued sample compared to that in the fresh one. This field-assisted rejuvenated behavior resembles that of (Bi,Pr)(Fe,Mn)$O_3$, in which the fatigue degradation was ascribed to the domain wall pinning by charged defects, such as electrons or $V_O$, while the microscopic studies of imperfections accompanied with the domain reversal was missing.[24]

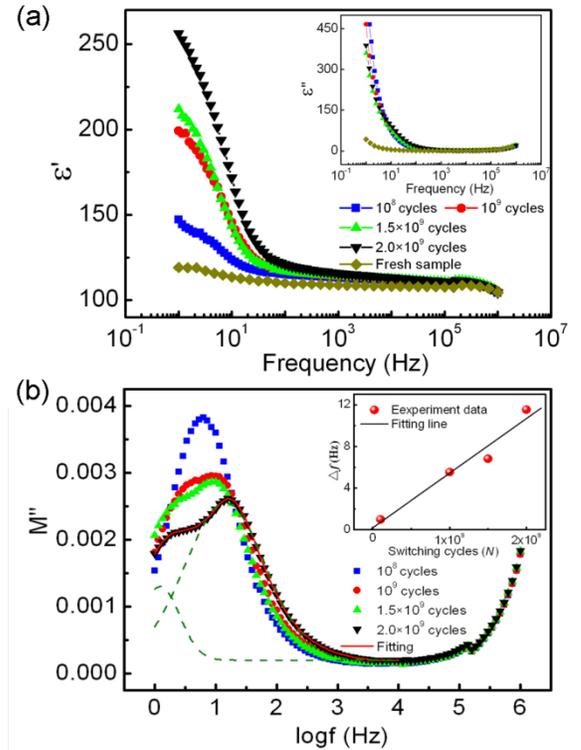

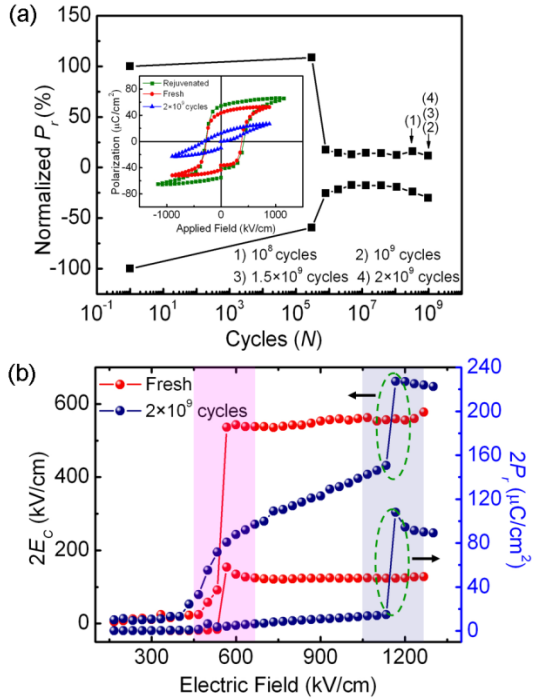

Figure 1. (a) Fatigue behavior of BFO thin film measured at room temperature, where four different switching stages are selected and marked as 1), 2), 3), and 4), respectively. Inset shows the P-E loops for the respective rejuvenated, fresh and fatigued BFO thin films. (b) Coecive field and remnant polarization as a function of

Figure 2. Variation of $\varepsilon'$ (a) and $M''$ (b) as a function of frequency for the BFO thin film measured at different stages of polarization switching. The inset of (a) shows frequency-dependency of dielectric loss ($\varepsilon''$). Inset of (b) shows the discrepancy ($\Delta f$) between frequencies of the separated peaks as a function of switching cycles (N).

**Defects Evolution during the Polarization Switching Process.** Figure 2a shows the frequency-dependent dielectric permittivity ($\varepsilon'$) for the BFO thin film, which was subjected to different numbers of switching cycles. For the fresh one, the $\varepsilon'$ remains largely unchanged, demonstrating an independent nature against the scanning frequency. With increasing switching cycles, a significant enhancement in $\varepsilon'$ was observed in the fatigued samples in the low frequency region (below 10 Hz), which is ascribed to the dielectric responses of space charges to the applied field.[25] In the case of the frequency domain, the dielectric response could be fully characterized by the complex permittivity ($\varepsilon^*$). However, the analysis of relaxation behavior is generally difficult to be achieved because of the significant contribution from *dc*-conductivity, which may obscure the relaxation peaks, especially in the low frequency region as indicated in inset of Figure 2a. An alternative approach to determine the exact position of characteristic peak is by the inversion method.[26] The electric modulus ($M^*$) can be used as a parameter to describe the dielectric response of non-conducting materials, which is a reciprocal of $\varepsilon^*$ and can be defined as follows:

$$M^* = M' + jM'' = \frac{1}{\varepsilon^*} = \frac{\varepsilon'}{|\varepsilon|^2} + j\frac{\varepsilon''}{|\varepsilon|^2} \quad (1)$$

where $M'$ and $M''$ are respective real and imaginary parts of $M^*$. The contribution from conductivity is therefore suppressed and the response of certain defects under *ac* field can be clarified in the $M''$ spectra as shown in Figure 2b, where an apparent change is observed at different stages of the fatigue process. A single relaxation peak at the initial fatigue stage is gradually evolved into partially overlapped double peaks, which tend to separate with increasing switching cycles and the peaks gap is enhanced as evidently shown in the inset of Figure 2b. The charge transport associated with a particular type of defect in dielectrics often exhibits a characteristic relaxation time ($\tau_r$), which is described in the frequency domain with $\tau_r = \frac{1}{2\pi f_p}$ and provides a direct pathway to disclose different dynamical processes, where $f_p$ is the corresponding peak frequency collected from the impedance spectra.[13] Therefore, the change in $M''$ spectra clearly indicates a dynamical evolution of charged defects during the fatigue process.

To clarify the role of defects involved during the cycling, the BFO thin films subjected to different numbers of switching cycles (labeled as sample_fresh, sample_$10^8$ and sample_$2\times10^9$) were selected to investigate the relaxation behavior by monitoring their temperature-dependent impedance spectra. The $M''$ spectra obtained from the sample_fresh are shown in Figure 3a. A single relaxation peak is observed and shifts toward the high frequency region with increasing temperature as indicative of thermal activated process. Identification of defect types involved in the sample_fresh was obtained by estimating the activation energy using the Arrhenius law:

$$f_p = f_0 \exp(-\frac{E_a}{K_B T}) \quad (2)$$

where $f_0$ is the pre-exponential factor and $E_a$ is the activation energy and $T$ is absolute temperature. As shown in Figure 3d, the activation energy thus derived is ~0.9 eV, which is comparable to the 0.99 eV measured in Bi:SrTiO$_3$, indicating the second ionized oxygen vacancy ($V_O^{\bullet\bullet}$) as the dominant defect existing in the fresh sample.[27]

In the sample_$10^8$, the polarization is dramatically degraded and the corresponding $M''$ spectra is shown in Figure 3b. For this sample, a single relaxation peak is also observed, which however locates at higher frequency at a given temperature compared to that in sample_fresh (e.g., at room temperature, the relaxation peak locates below 0.1 Hz for the sample_fresh, while above 1 Hz for the sample_$10^8$). It infers that there is a variation in the charged defects including propagation of existing ones or creation of new ones during the fatigue process.[26] The activation energy of the defect in the sample_$10^8$ is estimated to be ~0.2 eV, which is however far below the value for the sample_fresh. Similar activation energy values have been reported and considered to originate from the following possibilities:[26,28] (i) electron hopping along the chain of Fe$^{3+}$-Fe$^{2+}$ with an activation energy of ~0.3 eV; and (ii) diffusion of $V_O$ within their clusters, as the binding energy between the two $V_O$ are believed to be 0.2 eV. Consequently, a differentiation cannot be tentatively made between electrons and $V_O$ with these limited understandings.

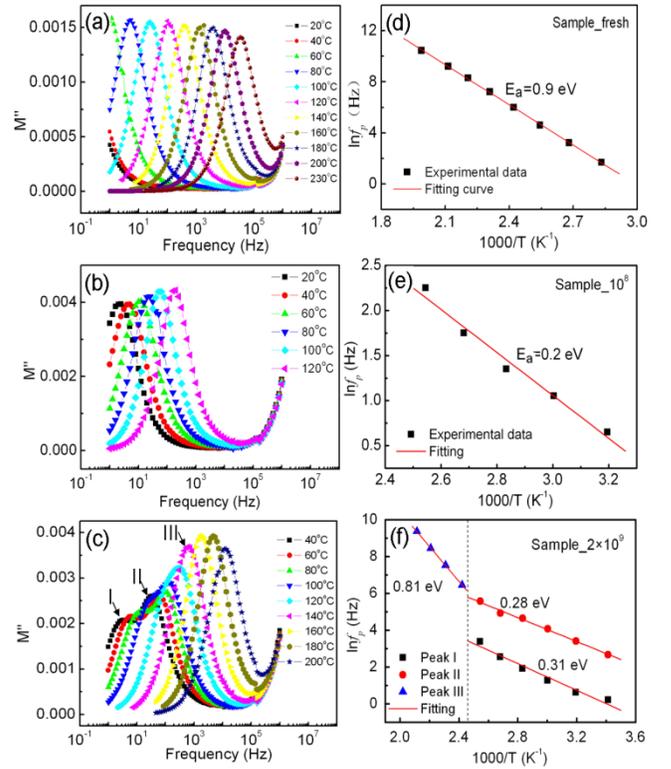

Figure 3. (a)-(c) The temperature-dependent $M''$ spectra of BFO thin film measured at different stages of polarization switching. (d)-(f) The corresponding Arrhenius plots of these samples.

Upon polarization switching for $2\times10^9$ cycles, the corresponding $M''$ spectra of sample_$2\times10^9$ is collected and shown in Figure 3c, which exhibits some new features. Below 120 °C, two partially overlapped peaks labeled I and II are observed in low frequency region (1-10 Hz) and higher frequency region (10-1000 Hz), respectively. One also notes that the peak I disperses much more strongly and approaches to peak II with increasing temperature. The partially overlapped two peaks appear to merge into one stronger peak (labeled as peak_III) when the temperature exceeds 120 °C. The three relaxation processes are closely related to the responses of defects and their respective natures are further identified by the Arrhenius plots shown in Figure 3f. The activation energies derived for these processes are 0.31, 0.28, 0.81 eV for peaks I, II, III respectively.

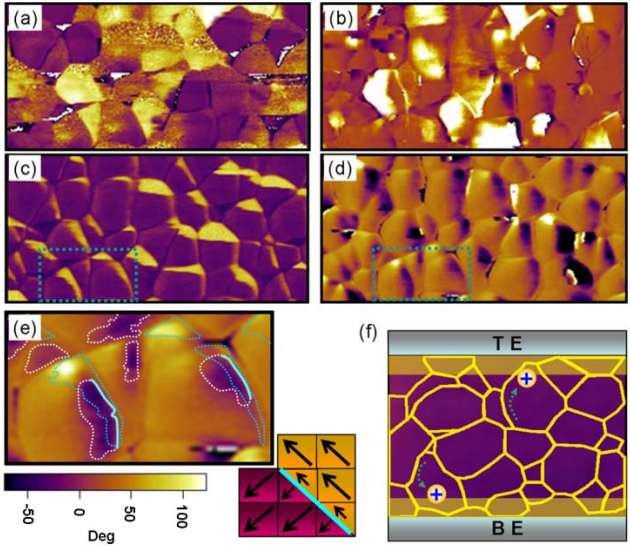

Figure 4. (a) OP and (b) IP PFM images (phase) for the as-deposited sample, (c) and (d) show the OP and IP PFM images (phase) of electrical-treated BFO thin film. (e) Zoomed-in IP PFM image of the green dotted square in (c) and (d). The blue and white dotted lines correspond to contrast between the bright and dark regions in OP and IP images, respectively. The sky-blue solid lines are the non-neutral domain walls, the structure of which is schematically shown in the bottom right corner image. (f) Schematic for the network-distributed $V_O$ along the framework of grain boundaries, where the movement of $V_O$ is facilitated.

For peak III, the value of 0.81 eV is consistent with the activation energy of 0.84 eV measured for the diffusion of oxygen ions in titanate-based bulk materials, such as $BaTiO_3$ and $Pb(Zr/Ti)O_3$.[29] Therefore, the relaxation process at temperatures above 120 °C can be unequivocally ascribed to the displacement of $V_O$. Nevertheless, in the aforementioned analysis, the relaxation process with activation energy around 0.2 eV may originate from the hopping of electrons or that of $V_O$ in clusters. Hereby, both possibilities have to be rationalized. Assuming that relaxation processes for peak I and II, which exhibit similar activation energies are due to electron hopping along the chain of $Fe^{3+}$-$Fe^{2+}$, one would see that the evolution of charged defects shifts from electrons- to $V_O$-driven mechanism at temperature of 120 °C. This assumption however is unreasonable, as the relaxation process reflected in the temperature-dependent impedance spectra is indeed of a thermal activated process, which should be varied in a consecutive way.[26,27,29] Generally, with increasing temperature, relaxation peak associated with a certain type of charged defect shifts gradually towards the high frequency side until moves out of the examining window instead of an "unexpected" absence. Therefore, the successive evolution of impedance spectra suggests that the origin for the relaxation processes of peak I and II should be related to the $V_O$ diffusion rather than the hopping of electrons.

Previous investigation conducted on $BaTiO_3$ has shown that $V_O$ can interact with each other and order themselves in a way to reach a stable structure. The binding energy for $V_O$ within the clusters was estimated to be ~0.2 eV, which is rather close to the experimental values obtained in the present work.[30] We therefore clarify that the hopping of $V_O$ between two alternative sites is responsible for the relaxation process of peak I and II. Moreover, one notes that the time constant $\tau_r$, which reflects the mobility of a certain type of charge defects responding to the external stimulus, can be used to distinguish the nature of charged defects.[13,31] Having carefully examined the results in Figure 3c, we note that peak frequencies of I and II are located below 10 Hz at room temperature. Therefore, the corresponding time constants are estimated to be >0.01 s, which is comparable to the values for $V_O$ reported in the previous work.[26] This evidence gives a strong support to the argument that the relaxation processes of peak I and II in sample_$2\times10^9$ originate from the migration of $V_O$.

Based on the above considerations, the evolution of relaxation peaks obtained for the sample_$2\times10^9$ is understandable. When temperature rises above 120 °C, isolated $V_O$ is released from the vacancy clusters through thermal dissociation, giving rise to the relaxation process of peak III. Moreover, we also note that the movement of an isolated $V_O$ in sample_$2\times10^9$ exhibits a lower activation energy of 0.81 eV compared to that of 0.9 eV in sample_fresh, indicating the propagation of $V_O$ during the fatigue process. The highly concentrated $V_O$ in the fatigued samples tend to be strongly inter-correlated with each other, and result in the formation of clusters to stabilize the lattice structure.

**Redistribution of Charged Defects within BFO Thin Films.** After clarifying the types of defects involved during fatigue process, here we investigate the redistribution of these impurities in the films. Complex impedance analysis was performed on BFO at varied fatigue stages. Figure S3 shows the Cole-Cole plot for the different fatigued samples in the modulus formalism, which can be clearly divided into triple semicircular arcs and are correspondingly attributed to the dielectric characteristics of grain, grain boundary, and electrode-film interfacial layer in a decreasing order of measuring frequencies respectively.[32,33] It is noted that the uniform semicircle in sample_$10^8$ tends to evolve into two separated semicircles with increasing switching cycles (Figure S3a). This evolution of the modulus planes clearly evidences that a largely different microstructure is formed between the grain boundary and interfacial layer upon electrical cycling. In the polycrystalline thin films, the impedance of grain boundary and interfacial layer is suggested to be dominated by trapped charges, such as impurities or defects in the samples.[34,35] Therefore, the fatigue-event-dependent formation of microstructure demonstrates that the occurrence of the redistributed defects is correlated with domain reversals.

The PFM method is employed to pole the sample to simulate the domain reversal and directly image the interplay between the configuration of microscopic domain structure and charged defects redistribution. The film surface is subjected to repeated write/read with a bias voltage of +/-10 V until the pinning domain image is evidently achieved. The out-plane (OP) and in-plane (IP) PFM images of the as-deposited sample are shown in Figure 4a,b, respectively, in which the piezoelectric polarities appear as yellow and purple regions randomly distributing throughout the film surface. We also note that some large grains within the fresh one are split into domains with opposite polarities, or adjacent parts of two neighboring grains merge into one domain, as indicative of a weak correlation between domains and crystallite structures.[36] Figure 4c shows the OP image of the BFO thin film after repetitive poling procedure, where one can see that the majority of piezoelectric polarities appear purple in color with the downward polarization, while there are some pinned areas showing opposite polarities in yellow appearance and locating in neighbor to the grain boundaries. The meandering image of domain contour agrees well with the domain wall pinning scenario, in which electrostatic coupling between a non-neutral domain wall and mobile carriers leads to the blocking of domain switching.

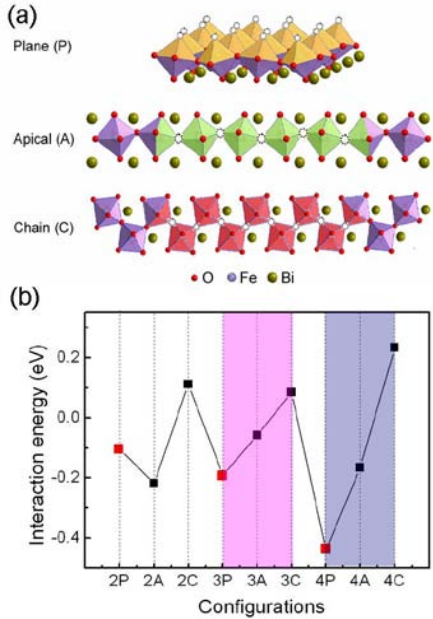

Figure 5. First-principles calculations for three types of ordered $V_O$ clusters: Plane, Apical, and Chain, which are represented by $n$P, $n$A, $n$C for cases with a loss of $n$ oxygen atoms (a) and the interaction energy (b) of the $V_O$ clusters. $V_O$ is shown by the dashed circle. The stabilities of clusters comprised of 3 and 4 $V_O$ are highlighted into pink and blue colors respectively, and the interaction energies of planar-aligned clusters are labeled in red squares.

The formation of pinning centers for domain walls can be achieved by the following two conditions:[17] (i) the presence of non-neutral domain walls; and (ii) inclusion of mobile carriers, which is common in oxide thin films, such as $V_O$ examined in our BFO sample. To properly elucidate domain wall pinning scenario, we conducted analysis on the configuration of domains and the domain walls in details. Figure 4e is the zoomed-in IP image of the area marked as the green dotted square in Figure 4c,d. The blue and white dotted lines correspond to the borderlines between the dark and bright regions in OP and IP PFM images respectively. The negatively charged domain walls are identified (marked into blue solid line in Figure 4e) and shown to be nearly meandering along the grain boundary regions. Charged domain walls were predicted with metallic behavior and shown to be correlated with the polarity of the majority carriers.[37] Therefore, the negatively charged domain apex occurs preferentially in the $p$-type systems in the view of Landauer instability and could be stabilized by the charged defects, including hole or $V_O$ to keep the local charge balance.

Charged domain walls in BFO thin films have been reported in the previous experimental studies. Zou *et al.* investigated the microscopic evolution of domain image during electrical cycling using PFM and Scanning Kelvin Probe Microscopy (SKPM).[6] They showed a direct observation of negatively charged domain walls during electrical cycling, which were over-compensated by $V_O$ to reduce the high electrical energy. In their work, the negative charged domain wall is parallel to the field direction in the epitaxially (001)-orientated BFO thin film and believed to play a minor role in fatigue degradation. In our polycrystalline thin films, the network-distributed grain boundaries, however, are deemed to block the movement of $V_O$ and serve as the preferential sites for $V_O$ to aggregate and form positive-charged clusters, which modulate the energy of domain apex by compensating the bound charges of domain walls and consequently reduce the wall's mobility.

Based on these understandings, the redistribution of charged defects during the fatigue process is unraveled: in sample_fresh, the dominant charged defects are identified to be isolated $V_O$, and the hopping of uniformly distributed $V_O$ within bulk interior is responsible for the microstructure response. With increasing switching cycles for $10^8$ cycles, the BFO thin film suffers from bad polarization degradation and the corresponding charged defects are shown to be correlated with $V_O$ clusters. We therefore rationalize the formation of microstructure in the sample_$10^8$ as a result of $V_O$ aligned along the grain boundary, which are network-distributed throughout bulk interior as schematically shown in Figure 4f. Upon polarization switching for $2\times10^9$ cylces, the polarization degradation arrives at a saturated stage, in which the domain reversal is largely suppressed. However, the movement of $V_O$ towards electrode stimulated by the cycling field can be facilitated through hopping between two alternative sites along the network-distributed channels and consequently results in the formation of an interfacial layer between electrode and BFO thin films.[38]

**First-principles Calculations.** We have identified that the unevenly distributed $V_O$ gives rise to the heterostructure across the BFO thin films, where the distortion of lattices is largely different. However, the stability and possible configuration of the nano-sized $V_O$ clusters within the BFO are still unclear, which play an important role in modulating switchable polarization by affecting the stress field and electronic structure of the thin films. To quantitatively estimate the stability and geometry of the ordered $V_O$ complex in the BFO thin film, we performed first-principles calculations based on density functional theory (DFT) to investigate the energetics of various clustering states of $V_O$. We studied three types of extended $V_O$ clusters (Figure 5): planes, apicals, and chains of $n$ vacancies denoted as $n$P, $n$C, and $n$A, respectively, which are viable in perovskite compounds. The interaction ener-gies ($E_{in}$) of these ordered vacancy clusters are calculated with the following formula:

$$E_{in}(nV_O) = E_{tot}(nV_O) + (n-1)E_{tot}(0V_O) - nE_{tot}(1V_O) \quad (3)$$

where $E_{tot}(nV_O)$ is the total energy of a cell containing $nV_O$. A negative sign of the $E_{in}(nV_O)$ indicates a stable configuration for the cluster whereas a positive sign suggests unstable binding. The variation of interaction energies of the defective structures with the loss of oxygen atoms up to four is plotted in Figure 5. Among all the ordered defects, oxygen divacancy is of utmost importance as divacancy serves as the seed nuclei for formation of larger clusters and its $E_{in}$ value, being in nature of binding energy, reflects an attractive or a repulsive trend between two isolated vacancies. The most stable binding configuration for a divacancy adopts an apical configuration with a binding energy of 0.22 eV, which is consistent with the present experimental results of the activation energy of $V_O$ estimated by the Arrhenius law and validates that clustering of $V_O$ is formed during the fatigue process. We also obtained a similar binding energy for the apical divacancy $V_O$ in a larger super cell containing 320 atoms indicating that the long-range relaxation of $V_O$ is small.

For higher ordered clusters (e.g., including 3 or 4 $V_O$ within the clusters), we find a relatively large exothermic process during the formation of $nP\text{-}V_O$ which suggests that two-dimensional planar self-organization of $V_O$ in BFO thin film is highly likely to occur. In contrast, the interaction energies of 2C, 3C, 4C-$V_O$ chains in BFO thin film are endothermic and chain-like-aligned $V_O$ are energetically unfavored, which however tend to jump and order themselves into planes to minimize the total energy. The size effect on the stability of plane-aligned $V_O$'s has also been monitored and the number of the contained $V_O$ is limited within four in this calculation because of the high computational demand. It shows that the interaction energy decreases with the increasing size of the planar $V_O$ cluster, indicating that possible stable states are arrived when the plane of $V_O$ is extended.

Although the defective states of isolated $V_O$ have been extensively studied,[39,40] the stability of single $V_O$ is less explored. Our study shows that clustering of $V_O$ is energetically favored in BFO thin film. Dynamic processes like melting and recombination of these ordered $V_O$ clusters occur with changing the temperature and the applied poling voltage until equilibrium is reached, which plays important roles in pinning the domain walls. Distortion of the cation sublattices and electrostatic field accompanied with these ordered vacancies may well be responsible for the degradation of the BFO based devices.

## CONCLUSIONS

In summary, we employ temperature-dependent impedance spectra in conjunction with PFM method to examine the evolution and redistribution of charged defects during the fatigue process. The fatigue failure of BFO thin films is shown to fall into domain-wall pinning scenario, where an increased number of $V_O$ engendered from the lattice distortion during electrical cycling tends to form extended clusters aligning along grain boundaries to block the domain reversal. First-principles calculations are employed to understand the stability of various ordered vacancies, which reveals that a planar self-organization of $V_O$ in BFO is energetically favored. Overall, this work establishes a linkage between the fatigue degradation and microscopic evolution of charged defects and clarifies the role of massively ordered $V_O$ clusters in the reduction of switchable polarization.

## MATERIASL AND METHODS

**Materials.** BFO thin films were deposited by RF magnetron sputtering on a $SrRuO_3$-buffered $Pt/TiO_2/SiO_2/Si$ substrate at temperature of 620 $^oC$. Firstly, a $SrRuO_3$ buffer layer of 80 nm was deposited on the substrate by RF sputtering. The BFO layer in thickness of 300 nm was subsequently deposited in the same system at a power of 120 W. For the deposition of both layers, the chamber pressure were kept at 1 mTorr with a Ar:$O_2$ ratio of 4:1. Prior to electrical measurements, Au dots of 200 μm in diameter were sputtered on the BFO film using a shallow mask to form top electrodes.

**Electrical Characterization.** The fatigue and ferroelectric behaviors of the BFO thin film was studied using a Radiant Precision Analyzer Station (Radiant Technologies, Medina, NY). During electrical cycling test, a bipolar triangle pulse with amplitude of 600 kV/cm at 100 kHz was used. The temperature-dependent impedance spectra studies were conducted using a Solartron Impedance Analyzer. The impedance measurements were carried out from room temperature to 230 °C in the frequency range of $1-10^6$ Hz.

**PFM Characterization.** PFM was carried out using a commercial AFM system (MFP-3 D, Asylum Research, USA ). Platinum coated silicon cantilever (radius of 15 nm with a spring constant of 2 N/m and a resonant frequency of 70 kHz) was used for the scanning with a tip lift height of 30 nm. AC signal with amplitude of 1 V at 100 kHz was applied between the tip (working as top electrode) and the bottom electrode of the sample to acquire the piezoresponse images of the out-plane and in-plane components. Through the same PFM tip, DC bias of $\pm 10$ V was applied locally to film surface of interest.

**Computational Details.** The first-principles periodic calculations based on DFT were performed using the Vienna Ab initio Simulation Package (VASP) within the framework of the GGA+U method.[41] The effective on-site Coulomb (U) parameter chosen for Fe-$3d$ orbitals was 4.0 eV.[42] A pseudocubic $2\sqrt{2} \times 2\sqrt{2} \times 4$ supercell (160 atoms) of rhombohedral $R3c$ phase was created to accommodate the G-type spin arrangement. The Brillouin-zone sampling was carried out using a $2 \times 2 \times 1$ Monkhorst-Pack $k$-point mesh. We employed the projected augmented wave (PAW) method and the wave functions were expanded in a plane-wave basis truncated at 500 eV. All the structures were relaxed until the Hellmann-Feynman forces become less than 0.01 eV/Å. Dipole-dipole interaction due to the periodic image of defective centers is considered for the calculation of the energetics of defective structures.